\documentclass[showpacs,amsmath,twocolumn,showkeys,pra]{revtex4-1}
\usepackage{dcolumn}    
\usepackage{bm}         
\usepackage{ifpdf}
\usepackage{amssymb,lineno,amsfonts}
\usepackage{graphicx}   
\usepackage{bbm}
\usepackage{mathrsfs}
\usepackage{upgreek}
\usepackage{mathtools}
\usepackage{epstopdf}
\usepackage{setspace}
\usepackage{hyperref}
\usepackage[matrix,frame,arrow]{xypic}
\usepackage{rotating}
\usepackage{float}
\usepackage{natbib}
\usepackage{color} 
\definecolor{med-blue}{RGB}{25,25,112} 
\hypersetup{colorlinks, linkcolor={red},citecolor={blue}, urlcolor={blue}}

\begin{document}

\vspace*{1cm}
\title{Benford's distribution in extrasolar world: Do the exoplanets follow Benford's distribution?}
\author{Abhishek Shukla$^1$, Ankit Kumar Pandey$^2$,  and Anirban Pathak$^1$}
\email{anirban.pathak@gmail.com}
\affiliation{$^1 $Jaypee Institute of Information  Technology, A-10, Sector 62, Noida, UP 201307,  India\\
$ ^2 $ Indian Institute of Science Education and Research, Mohali 411008, India\\}

\begin{abstract}
{In many real life situations, it is observed that the first digits (i.e., $1,2,\ldots,9$) of a numerical data-set, which is expressed using decimal system, do not follow a random distribution. Instead, the probability of occurrence of these digits decreases in almost exponential fashion starting from 30.1$\%$ for 1 to 4.6$\%$ for 9. Specifically, smaller numbers are favoured by nature in accordance with a logarithmic distribution law, which is referred to as Benford's law. The existence and applicability of this empirical law have been extensively studied by physicists, accountants, computer scientists, mathematicians, statisticians, etc., and it has been observed that a large number of data-sets related to diverse problems follow this distribution. However, except two recent works related to astronomy, applicability of Benford's law has not been tested for extrasolar objects. Motivated by this fact, this paper investigates the existence of Benford's distribution in the extrasolar world using Kepler data for exoplanets. The investigation has revealed the presence of Benford's distribution in various physical properties of these exoplanets. Further, Benford goodness parameters are computed to provide a quantitative measure of coincidence of real data with the ideal values obtained from Benford's distribution. The quantitative analysis and the plots have revealed that several physical parameters associated with the exoplanets (e.g., mass, volume, density, orbital semi-major axis, orbital period, and radial velocity) nicely follow Benford's distribution, whereas some physical parameters (e.g., total proper motion, stellar age and stellar distance) moderately follow the distribution, and some others (e.g., longitude, radius, and effective temperature) do not follow Benford's distribution. Further, some specific comments have been made on the possible generalizations of the obtained result, its potential applications in analyzing data-set of candidate exoplanets, and how interested readers can perform similar investigations on other interesting data-sets.}
\end{abstract}

\maketitle

\section{Introduction} \label{intro}
\begin{figure}
\begin{center}
\hspace{-1cm}
\includegraphics[trim=0cm 0cm 0cm 0cm, clip=true,width=9.5cm]{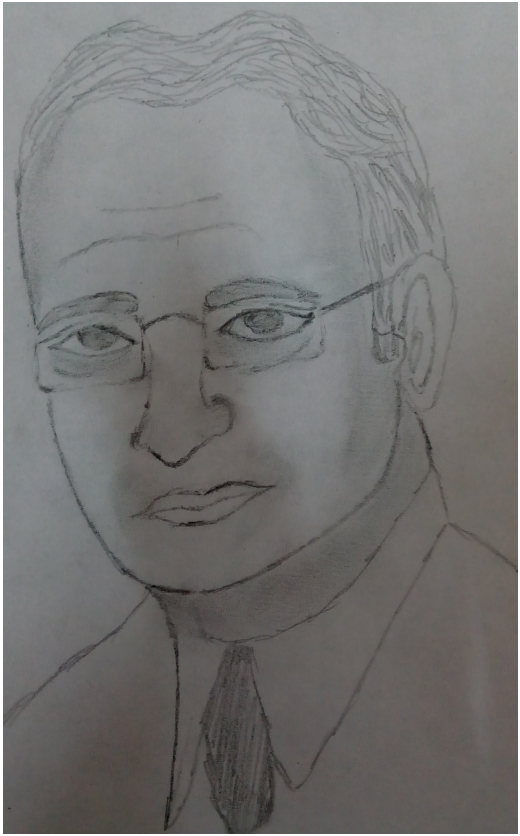} 
\caption{A pencil sketch of Frank Benford.}
\label{photo} 
\end{center}
\end{figure}

In 1881, while going through the logarithms of an unbiased data-set, Simon Newcomb noticed an anomalous behavior in the distribution of digits \cite{newcomb}. Actually, he computed occupancy of most significant digit (MSD) from such a data-set. Counter to common intuition, which would expect an unbiased or random behavior in occupancy of the digits, Newcomb found that it decreases exponentially with digits. The probability of occurrence of 1 was found to be 30.1 $\%$ for 1 and the same for 9 was found to be 4.6 $\%$. Simon's prediction was empirical in nature, and due to lack of mathematical structure his article did not receive much attention. Later in 1938, Benford, (see image shown in Fig. \ref{photo}) mathematically formulated a law to calculate probability $P_{d}$ of occurrence of  the digit $d$ as the  MSD, with the  sum of the probability to be unity (i.e., $\sum\limits_{1}^{9} P_{d}= 1$) \cite{benford}. The probability distribution introduced by Benford was
\begin{equation}
P_{d}=\log_{10} \left(1 + \frac{1}{d} \right). \label{benford1}
\end{equation}
%

Since the pioneering works of Newcomb and Benford, a large number of works related to Benford's law have been reported in various contexts  (for a fascinating history of Benford's law, see \cite{berger2011basic,adhikari1968distribution,benbiblio}). For example, it's presence and applicability have been investigated in  various domains, like astrophysics \cite{Astrophysical1,alexopoulos2014benford}, geography \cite{geographical}, biology \cite{neural,biology2,aerobiological}, seismography \cite{seismic}, stock market and accounting \cite{stockmarket,financial1}. Interestingly, violation of Benford's law has been found to be capable of  detecting cases of tax fraud \cite{taxfruad} and election fraud \cite{electionfraud}, and it's routinely used by accounting professionals to detect financial irregularities. However,  its  reliability as a tool for fraud detection is still debatable. The issues and the cases where  violation  of Benford's law do not correctly predict presence of fraud are discussed in \cite{taxfruad}. It is not our purpose to discuss this particularly interesting issue in detail, rather we are interested to note that Benford's law-based analysis has recently drawn considerable attention of physics community. Especially, after it's formulation as an efficient tool to study quantum phase transitions by  De and Sen  \cite{ujjwal1} and Rane et al.,  \cite{ujjwal2}. Further, it has also been shown that Benford's law-based analysis is helpful in spectroscopy. In particular, its applications for weak peak detection, phase correction, and  baseline  correction have been demonstrated on NMR signal by some of the present authors \cite{bhole2015benford}. A good agreement with ideal Benford's distribution was observed  in various NMR spectra, and that validated the existence of Benford's distribution in NMR-based systems. Furthermore, an attempt to use Benford law-based analysis for processing MRI data has already been made in Ref. \cite{bhole2015benford}. This provides us an excellent example of application of Benford's law. In addition, in an attempt to reveal the existence of first-principle-based rule behind Benford's law, Shao et al., have reported Benford's distribution for various statistical ensembles \cite{shao2010significant}. They found that Maxwell-Boltzmann and Bose-Einstein statistics allow periodic fluctuations in occupancy of digits with temperature, while for Fermi-Dirac statistics such fluctuations remains absent \cite{shao2010significant}.

Until recent past, all the investigations on the Benford's law were restricted to the data-set generated in context of our solar system in general, and the Earth in particular. However, recent astrophysical observations reported in \cite{Astrophysical1} and \cite{alexopoulos2014benford} have established that Benford's law is followed by star distances and distances from the Earth to galaxies. This observation, and the fact that one of the most prominent interest of mankind is to find promising sites to host an extra-terrestrial life, have motivated us to ask: "Do exoplanets follow Benford's distribution?" We try to answer this particular question using Kepler data \cite{keplerarchive}, which provides various information related to exoplanets that are mainly detected by NASA's Kepler telescope. Size of this data-set (i.e., Kepler data) has been considerably increased recently as NASA has confirmed the existence of several exoplanets. With this new announcement, the number of detected and confirmed exoplanets  goes to $\approx$ 3300. Which is a reasonable size for statistical analysis of the data-set in general, and for investigation on the existence of Benford's distribution, in particular. This point would be more clear if we note that in  Ref. \cite{pintr2013statistical,pintr2014relative}, the statistical analysis of exoplanets data was performed by some of the present authors using a data-set of (1771) exoplants, which was the number of exoplanets known at that time. Still, Ref. \cite{pintr2014relative} yielded various interesting results related to the possibility of existence of habitable exoplanets \cite{natureasia}. \color{black}

Remaining part of this paper is organized as follows. In Sec. \ref{mth}, we briefly describe the method adopted here for the investigation of  Benford's distribution, and the method adopted for computing Benford goodness parameter (BGP), which may be viewed as a measure of similarity between the Benford's distribution, and the actual distribution. In Sec. \ref{re}, we describe our results. Finally, we conclude the paper in Sec. \ref{conc}, where we have also mentioned some potential applications of the present work.

\section{Method} \label{mth}
 To calculate Benford's distribution for a given data-set we have adopted the simplest method described in Ref. \cite{benfordonexcel}, where it is shown that the distribution of MSD can be obtained using a spread-sheet (Microsoft Excel or a similar program). Using the above mentioned procedure we have calculated probability of occurrence for each digit in the Kepler data for exoplanets \cite{keplerarchive}.  The same is illustrated through a set of plots. Specifically, in Fig. \ref{survey1} we illustrate the  distribution of MSDs for
 Kepler data for exoplanets. All sub-plots of Fig. \ref{survey1}, clearly show that the values of a set of physical properties (e.g., mass, volume, density, orbital semi-major axis, orbital period, and radial velocity) are distributed in a manner that nicely matches with Benford's distribution. In other words, Fig. \ref{survey1} shows that exoplanets follow Benford's distribution. However, in all the subplots, the matching between the real distribution and  the ideal Benford's distribution is not the same. Thus, to understand how closely the values associated with a particular property follow Benford's distribution, we need a quantitative measure. Interestingly, such a quantitative measure exists \cite{gaurav}, and referred to as the BGP. For a given  data-set, BGP is defined as  
\begin{widetext} 
 \begin{eqnarray}
 \rm{BGP}=\Delta P = 100\left( 1 - \sqrt{ \sum_{d=1}^{9} \frac{(P(d)-P_B(d))^2}{P_B(d)}} ~\right).
 \end{eqnarray} 
\end{widetext}
 Here, $P(d)$ is the observed probability for digit $d$ and $P_{B}(d)$ is the ideal probability in Benford's distribution for the same digit $d$.  A complete overlap corresponds to $\Delta P=100$, but there is no lower limit. Thus, the larger the value of $\Delta P$ or BGP for a data-set, the closer it is to the ideal Benford's distribution. In other words, we may use $\Delta P$ or BGP as a quantitative measure of how accurately a given data-set follows Benford's distribution. In the following section we have used this measure to analyze Kepler data for exoplanets \cite{keplerarchive}. Specifically, data for density, orbital period, and orbital semi-major axis were taken from \cite{keplerarchive} on May 17, 2016, while data for the  rest of the quantities have been taken from the same data archive on May 04, 2016.
 
 Driven by the curiosity of examining deeper statistical symmetry present in Kepler data for exoplanets, we have also calculated joint probability of occupancy $P(d_1,d_2)$ for first and second significant digits being $d_1$ and $d_2$, respectively. For the purpose, we have used Hill's \cite{hill1995base} formula for generalized Benford distribution, which states that the probability $P(d_1,d_2, \cdots d_N)$ for digits $d_1$, $d_2$, $\cdots$ $d_N$ is
 
 \begin{equation}
 P(d_1,d_2, \cdots d_N) = \mathrm{log_{10}}\left[1 + \left(\sum_{i=1}^{k} d_k 10^{k-i}\right)^{-1}\right].  \label{benford2}
 \end{equation}
In particular, using  Eq. \ref{benford2},  we have calculated $P(d_1,d_2)$ for some quantities namely mass, volume, orbital period, effective teamprature and radius. We found encouraging results in case of orbital period, but not in other cases. In next section we will discuss these results in detail.
 \color{black}
\section{Result and Discussion} \label{re}
\begin{figure}[h]
\begin{center}
\includegraphics[trim=0cm 0cm 0cm 0cm, clip=true,width=8.8cm]{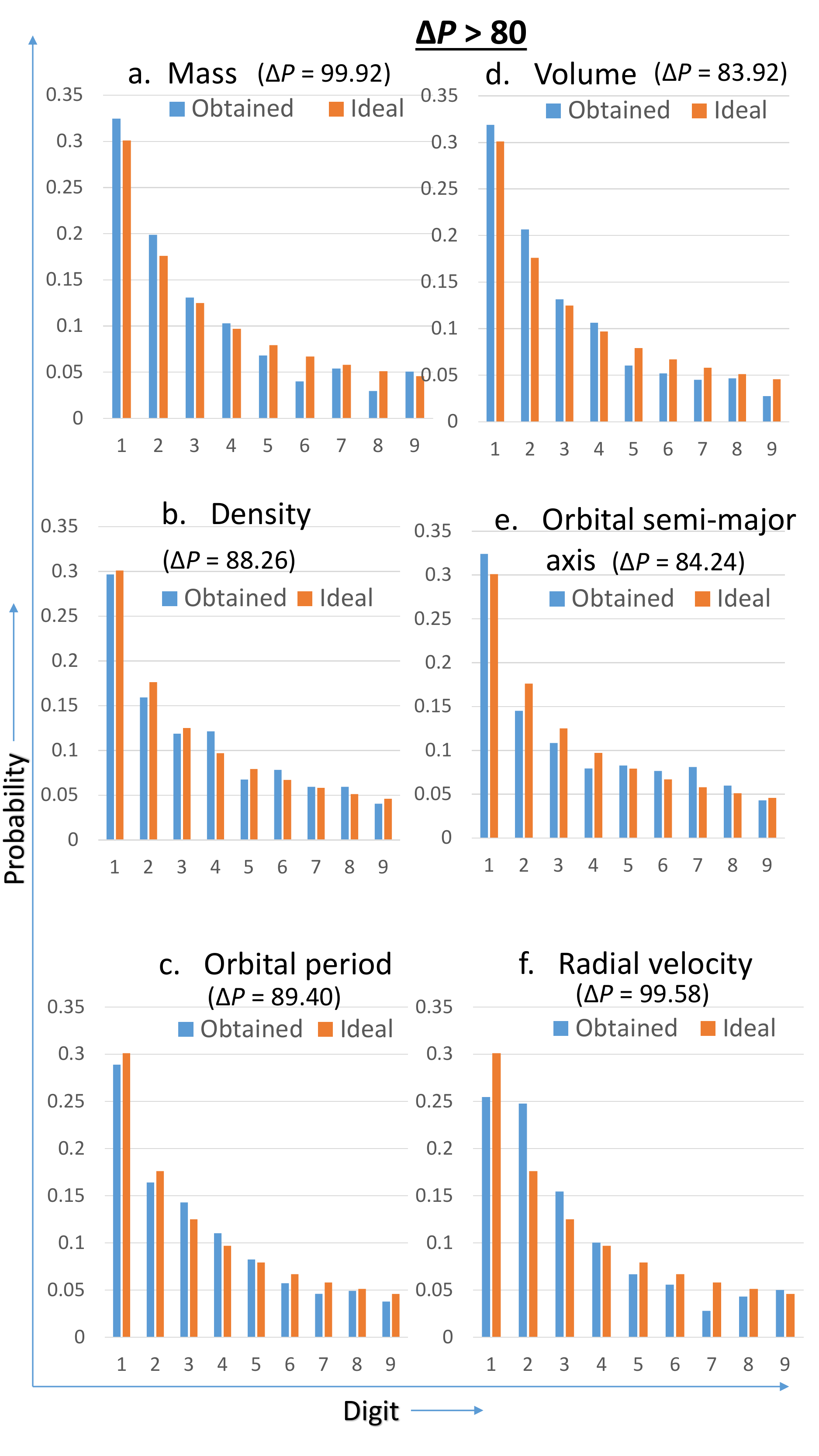} 
\caption{(Color online) 
Figure contains probability (vertical axis) of occurrence of MSD (horizontal axis). Ideal Benford's distribution is shown in orange color and the observed probability distributions are  shown in Blue color.  Subplots (a-f) show Benford's distributions of MSD for various physical quantities obtained from the Kepler archive \cite{keplerarchive}. In each subplot corresponding BGP value is noted.}
\label{survey1} 
\end{center}
\end{figure}

\begin{figure}[h]
\begin{center}
\includegraphics[trim=0cm 0cm 0cm 0cm, clip=true,width=8.8cm]{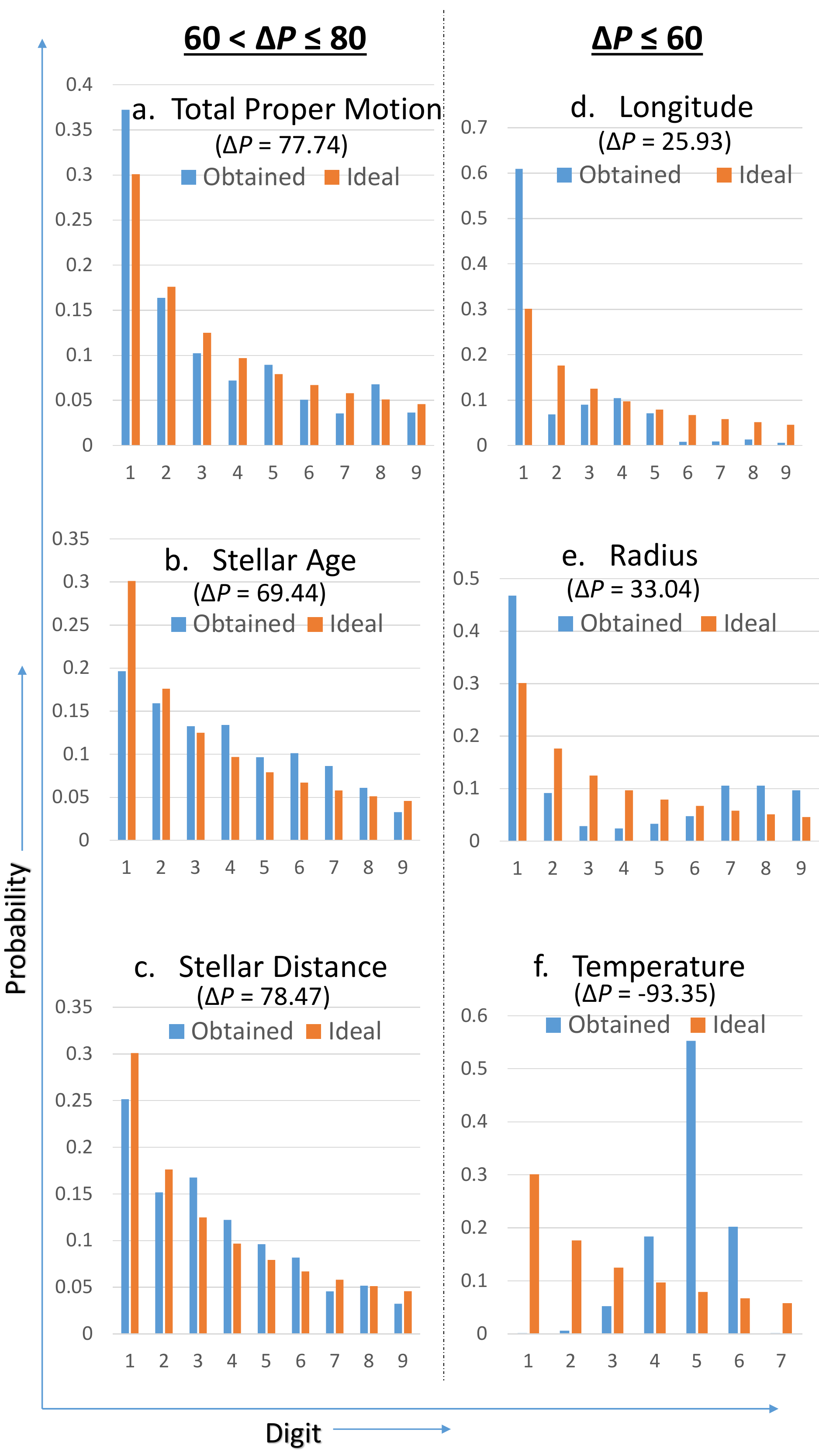} 
\caption{(Color online) Figure contains probability (vertical axis) of occurrence vs MSD (horizontal axis). Similar to previous figure, ideal Benford's distribution is shown in orange color and the observed probability distribution is shown in Blue color. Left column contains those physical quantities who have BGP, $60<$ $\Delta$ P $\leq{80}$, whereas, right column incorporates those physical quantities who have BGP $\le{60}$. The physical quantity and the corresponding BGP value of the distribution are noted in every subplot.}
\label{survey2} 
\end{center}
\end{figure}

Fig. \ref{survey1} illustrates plots for observed and ideal Benford's distribution  for various physical quantities, namely mass, density, orbital period, volume, orbital semi-major axis, and radial velocity of exoplanets with BGP values 99.92, 83.92, 89.40, 83.92, 84.24, 99.58, respectively. 
From  Fig. \ref{survey1}, we find that mass of exoplanets most closely follows Benford's distribution as BGP for this set of data is 99.92 (cf. Fig. \ref{survey1} a). Here, it may be noted that in Fig. \ref{survey1} a, the probabilities of occurrence of the MSDs are computed  after multiplying Jupiter mass into the data obtained from the Kepler archive. Similarly, in Fig. \ref{survey2} d, 
radius used for calculating volume is absolute, as it has been obtained by multiplying  the radius of the Earth into the data obtained from the Kepler archive. However, these scalings have not effected the distribution of MSDs, as Benford's distribution is known to be scale-independent. Further, from  Fig. \ref{survey1} we observe that the exoplanets' mass (cf. Fig. \ref{survey1} a), density  (cf. Fig. \ref{survey1} b), orbital period (cf. Fig. \ref{survey1} c),  volume (cf. Fig. \ref{survey1} d), orbital semi-major axis (cf. Fig. \ref{survey1} e), and radial velocity (cf. Fig. \ref{survey1} f) nicely follow Benford's distribution. Specifically, BGP computed for all these physical properties are greater than 80.  However, the Kepler data for other physical properties don't follow Benford's law so strictly. To be precise,  we have observed that there are some quantities, which have only moderate overlap with the ideal Benford's distribution. These quantities are total proper motion, stellar age and stellar distance of exoplanets. In fact, quantitative measure of BGP allows us to construct a criteria for classifying data depending on their BGP values. Specifically, we consider that BGP values greater than 80 correspond to a good agreement, BGP values in the range ($60<\rm{BGP}\le80$) correspond to moderate or intermediate agreement, and BGP values  $\leq60$ implies that data-set don't follow Benford's distribution or equivalently, bad agreement.
Now, we may look at  Fig. \ref{survey2}, which contains observed and ideal Benford's distribution plots for various physical quantities that are not illustrated in the previous plot.  Fig. \ref{survey2} (a-c)  illustrate distribution of first digits for all those quantities that moderately  follow Benford's distribution (thus, their BGP values are in  the  range $60<\rm{BGP}\le 80$). This set includes total proper motion, stellar age, and stellar distances, and corresponding BGP values are 77.74, 69.44, and 78.47, respectively. Similarly,  Fig. \ref{survey2} (d-f) illustrate the statistical distribution of most significant  digits for values of longitude (in radians), radius, and effective temperature \cite{wikiefftemp}, and it is observed that data-set for these properties do not follow Benford's distribution. The same is also quantitatively reflected in the BGP values  (BGP $\le 60$) obtained for longitude, radius and effective temperature. On keen observation, we noticed that those quantities who do not have good BGP values were actually having small variation in data, typically they are of the same order (or variation in orders of magnitude is very narrow)  and hence have lower BGP. Such a data-set  may be considered as biased data-set. This explains, why Benford like distribution is not observed in Fig.  \ref{survey2} (d-f) (i.e., for longitude, radius and effective temperature of exoplants). Thus, in brief, we may state that leaving a few incidents of biased data, it is observed in general that Benford's distribution is followed by the values of most physical properties associated with the exoplanets.

\begin{widetext}

 \begin{figure}
 \begin{center}
 \includegraphics[trim=0cm 0cm 0cm 0cm, clip=true,width=15cm]{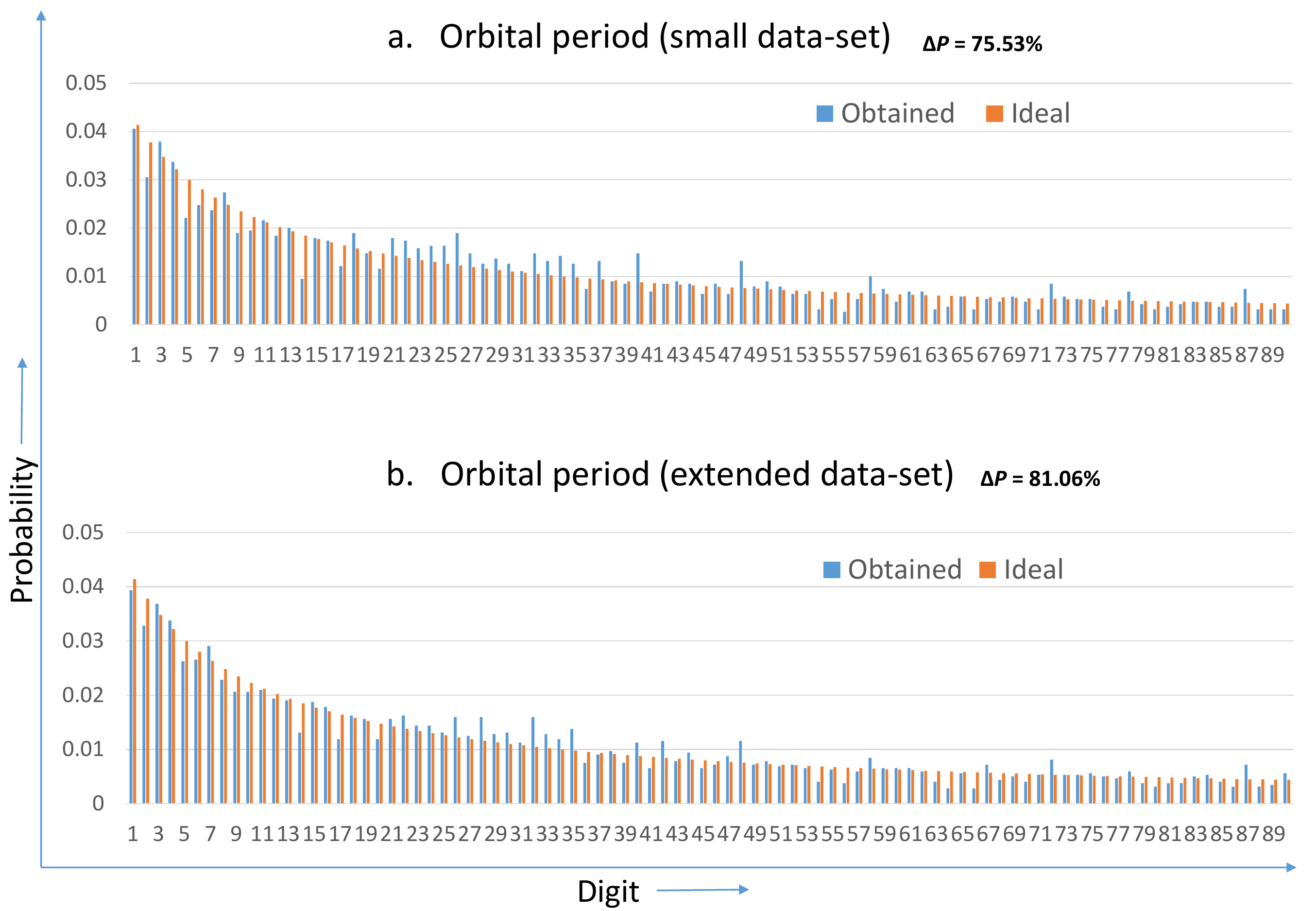} 
 \caption{(Color online) Figure contains probability (vertical axis) vs first and second significant digits $d_1d_2$ (horizontal axis). Again ideal Benford's distribution is shown in orange color and observed probability distribution is shown in blue color. Upper panel shows results on a  data-set of 1898 exoplanets while lower panel illustrates the results for an enlarged data-set of 3207 exoplanets. Corresponding BGPs are also mentioned  (75.53 for 1898 exoplanets and 81.06 for 3207 exoplanets).}
 \label{survey3} 
 \end{center}
 \end{figure}
\end{widetext}

Inspired by the observation that MSDs for values of various properties associated with exoplanets follow Benford's distribution, we tried to to investigate whether the second MSDs also follow this distribution. To do so, we have computed  $P(d_1,d_2)$  using Eq. \ref{benford2} for a few physical properties (e.g., orbital period, mass and volume). Here, we illustrate our observations on orbital period only. In Fig. \ref{survey3}, the overlap of $P(d_1,d_2)$ obtained from real data and ideal Benford's distribution is shown. It's observed that BGP increases with the size of data-set. In particular, for a data-set of 1898 exoplanets, we obtained BGP= 75.53, and for a data-set of 3207 exoplanets we obtained BGP=81.06. Surprisingly, we did not observe this increase in BGP with size of data-set for other physical properties (mass and volume). 
A probable reason for this observed increase in BGP of two digit distribution $P(d_1,d_2)$ for orbital period, but not in the case of other properties may be that for the former case data-size is sufficiently large to yield unbiased nature of data so that it can follow two digit Benford distribution and hence becomes close to ideal Benford distribution. In contrast,  for other properties, may be the data-size is still not sufficient for realization of unbiased two digit data-set.
\color{black}

\section{Conclusions} \label{conc}
The validity of Benford's law is investigated for the first time for exoplanets. The investigation is performed using Kepler data, and  it is observed that the data-set corresponding to exolanets mass, volume, density, orbital semi-major axis, orbital period and radial velocity nicely follow this law, whereas exoplanets' total proper motion, stellar age, and stellar distance moderately follow Benford's law, and exoplanets' longitude, radius, and effective temperature hardly follow the law. This is illustrated through Figs. \ref{survey1} - \ref{survey3}, and clearly established by a quantitative measure called BGP. It is found that the BGP is highest for  the  mass of the exoplanets (99.92), and it is extremely high for  the  radial velocity of the exoplanets (99.58), too. Thus, these two parameters almost exactly follow Benford's distribution as the upper-bound for BGP is 100. Such a high BGP is rare for any data-set, and this observation has provided us a clear affirmative answer to the question that we asked in the beginning: Do the exoplanets follow Benford's distribution?

Statistical distribution of exoplanets' mass and radial velocity clearly established the fact that exoplanets  follow Benford's distribution, and validity of Benford's law is not restricted to earth, rather it's universal as it's followed in extrasolar worlds, too. This strong observation is further supported by the fact that exoplanets' density (BGP=88.26), orbital periods (BGP=89.40), volume (BGP=83.92), and orbital semi-major axis (BGP=84.24) also strongly follow Benford's law. Thus, this empirical law seems to be universal and probably it's more fundamental and profound in nature than it's understood to be. However, some questions are still open. It's not well understood (physically), why it works for certain data-set, and why it does not work for others. What is obtained until now is a mathematical insight that helps us to understand where (i.e., in which data-sets) it works and where it does not. For example, we know that the a data-set which is not biased and where the order of magnitude varies considerably, is expected to follow this law. This point only answers "where" or "when", but neither provides any physical insight (an understanding from the first principle) nor answer "why".  Thus, it needs more investigations. Interestingly, this type of investigation does not require sophisticated equipments or software. One can just use a spread-sheet and follow the steps given in Ref. \cite{benfordonexcel} to check the existence of Benford's distribution in other data-set. One may also generalize the results easily. For example, one may examine whether the Benford's distribution is followed by second, third, fourth,... significant digits by using a general version of Benford's distribution introduced by Hill \cite{hill1995base}, 

%

 Recently, Eq. \ref{benford2} has been used in Ref.  \cite{alexopoulos2014benford} to establish that MSD for star distances  agrees  very well with the Benford's distribution as far as the first, second and third significant digits are concerned. Similar exercises can be performed using Kepler data and other available data-sets of interest. Motivated by this fact, we have computed $P(d_{1},d_{2})$ using Eq. \ref{benford2} for a set of physical properties. Another version of Benford's law has been introduced in \cite{miller2015theory}, and the authors have referred to it as strong Benford's law, which provides the probability of a specific number "$s$" in a data-set ( see definition 1.4.3 in \cite{miller2015theory} ). One can also check validity of strong Benford's law for a data-set using a spread-sheet. Thus, the presented result can be generalized, and similar results can be obtained in other data-sets of interest. However, before we perform such an exercise, we must ask whether it is worthy to perform such an exercise? Whether such an investigation is expected to provide some physical information or new insights to the data-set. The answer is yes, and in what follows, we elaborate this by discussing a specific possibility.

We have already mentioned that Benford's distribution has been successfully used in accounting to detect frauds \cite{taxfruad}, which may be viewed as a noise introduced by a person or a group of person in a data-set which was otherwise expected to follow Benford's distribution. Now,  the present paper establishes that several physical parameters associated with the exoplanets nicely follow Benford's distribution. This implies that in analogy with accounting, we may try to locate noise (which is analogous to fraud in accounting) in the Kepler data-set of candidate exoplanets (a set of potential exoplanets whose status are not yet confirmed), and that can ease our effort to locate actual exoplanets.

 Finally, we would like to note that statistical analysis of Kepler data is not new. Earlier studies performed by some of us \cite{pintr2014relative}  had revealed the region, where to look for habitable exoplanets, and the present study hints for a method to analyze candidate exoplanets. Keeping all these in mind, we conclude the paper optimistically, with a hope that this work would lead to a few more statistical investigations in the similar directions and those investigations would provide more physical insights on: Why does Benford's law work universally?

\textbf{Acknowledgment:}

 AP and AS thank Defense Research $\&$ Development Organization (DRDO), India for the support provided through the project number ERIP/ER/1403163/M/01/1603. AKP thanks JIIT, Noida (where this work is done) for the hospitality and facilities provided during his visit as a summer intern. 

\bibliographystyle{apsrev4-1}
\bibliography{ref}

\begin{thebibliography}{29}%
\makeatletter
\providecommand \@ifxundefined [1]{%
 \@ifx{#1\undefined}
}%
\providecommand \@ifnum [1]{%
 \ifnum #1\expandafter \@firstoftwo
 \else \expandafter \@secondoftwo
 \fi
}%
\providecommand \@ifx [1]{%
 \ifx #1\expandafter \@firstoftwo
 \else \expandafter \@secondoftwo
 \fi
}%
\providecommand \natexlab [1]{#1}%
\providecommand \enquote  [1]{``#1''}%
\providecommand \bibnamefont  [1]{#1}%
\providecommand \bibfnamefont [1]{#1}%
\providecommand \citenamefont [1]{#1}%
\providecommand \href@noop [0]{\@secondoftwo}%
\providecommand \href [0]{\begingroup \@sanitize@url \@href}%
\providecommand \@href[1]{\@@startlink{#1}\@@href}%
\providecommand \@@href[1]{\endgroup#1\@@endlink}%
\providecommand \@sanitize@url [0]{\catcode `\\12\catcode `\$12\catcode
  `\&12\catcode `\#12\catcode `\^12\catcode `\_12\catcode `\%12\relax}%
\providecommand \@@startlink[1]{}%
\providecommand \@@endlink[0]{}%
\providecommand \url  [0]{\begingroup\@sanitize@url \@url }%
\providecommand \@url [1]{\endgroup\@href {#1}{\urlprefix }}%
\providecommand \urlprefix  [0]{URL }%
\providecommand \Eprint [0]{\href }%
\providecommand \doibase [0]{http://dx.doi.org/}%
\providecommand \selectlanguage [0]{\@gobble}%
\providecommand \bibinfo  [0]{\@secondoftwo}%
\providecommand \bibfield  [0]{\@secondoftwo}%
\providecommand \translation [1]{[#1]}%
\providecommand \BibitemOpen [0]{}%
\providecommand \bibitemStop [0]{}%
\providecommand \bibitemNoStop [0]{.\EOS\space}%
\providecommand \EOS [0]{\spacefactor3000\relax}%
\providecommand \BibitemShut  [1]{\csname bibitem#1\endcsname}%
\let\auto@bib@innerbib\@empty
\bibitem [{\citenamefont {Newcomb}(1881)}]{newcomb}%
  \BibitemOpen
  \bibfield  {author} {\bibinfo {author} {\bibfnamefont {S.}~\bibnamefont
  {Newcomb}},\ }\href@noop {} {\bibfield  {journal} {\bibinfo  {journal} {A. J.
  of Math.}\ }\textbf {\bibinfo {volume} {4}},\ \bibinfo {pages} {39} (\bibinfo
  {year} {1881})}\BibitemShut {NoStop}%
\bibitem [{\citenamefont {Benford}(1938)}]{benford}%
  \BibitemOpen
  \bibfield  {author} {\bibinfo {author} {\bibfnamefont {F.}~\bibnamefont
  {Benford}},\ }\href {http://www.jstor.org/stable/984802} {\bibfield
  {journal} {\bibinfo  {journal} {Proc. of A. phil. Society}\ }\textbf
  {\bibinfo {volume} {78}},\ \bibinfo {pages} {551} (\bibinfo {year}
  {1938})}\BibitemShut {NoStop}%
\bibitem [{\citenamefont {Berger}\ \emph {et~al.}(2011)\citenamefont {Berger},
  \citenamefont {Hill} \emph {et~al.}}]{berger2011basic}%
  \BibitemOpen
  \bibfield  {author} {\bibinfo {author} {\bibfnamefont {A.}~\bibnamefont
  {Berger}}, \bibinfo {author} {\bibfnamefont {T.~P.}\ \bibnamefont {Hill}},
  \emph {et~al.},\ }\href@noop {} {\bibfield  {journal} {\bibinfo  {journal}
  {Probability Surveys}\ }\textbf {\bibinfo {volume} {8}},\ \bibinfo {pages}
  {1} (\bibinfo {year} {2011})}\BibitemShut {NoStop}%
\bibitem [{\citenamefont {Adhikari}\ and\ \citenamefont
  {Sarkar}(1968)}]{adhikari1968distribution}%
  \BibitemOpen
  \bibfield  {author} {\bibinfo {author} {\bibfnamefont {A.}~\bibnamefont
  {Adhikari}}\ and\ \bibinfo {author} {\bibfnamefont {B.}~\bibnamefont
  {Sarkar}},\ }\href@noop {} {\bibfield  {journal} {\bibinfo  {journal}
  {Sankhy{\=a}: The Indian Journal of Statistics, Series B}\ ,\ \bibinfo
  {pages} {47}} (\bibinfo {year} {1968})}\BibitemShut {NoStop}%
\bibitem [{\citenamefont {Berger}\ \emph {et~al.}(2009)\citenamefont {Berger},
  \citenamefont {Hill},\ and\ \citenamefont {Rogers}}]{benbiblio}%
  \BibitemOpen
  \bibfield  {author} {\bibinfo {author} {\bibfnamefont {A.}~\bibnamefont
  {Berger}}, \bibinfo {author} {\bibfnamefont {T.~P.}\ \bibnamefont {Hill}}, \
  and\ \bibinfo {author} {\bibfnamefont {E.}~\bibnamefont {Rogers}},\
  }\href@noop {} {\enquote {\bibinfo {title} {Benford online bibliography},}\
  }\bibinfo {howpublished}
  {\url{http://www.benfordonline.net/list/chronological}} (\bibinfo {year}
  {2009})\BibitemShut {NoStop}%
\bibitem [{\citenamefont {Moret}\ \emph {et~al.}(2006)\citenamefont {Moret},
  \citenamefont {de~Senna}, \citenamefont {Pereira},\ and\ \citenamefont
  {Zebende}}]{Astrophysical1}%
  \BibitemOpen
  \bibfield  {author} {\bibinfo {author} {\bibfnamefont {M.~A.}\ \bibnamefont
  {Moret}}, \bibinfo {author} {\bibfnamefont {V.}~\bibnamefont {de~Senna}},
  \bibinfo {author} {\bibfnamefont {M.~G.}\ \bibnamefont {Pereira}}, \ and\
  \bibinfo {author} {\bibfnamefont {G.~F.}\ \bibnamefont {Zebende}},\
  }\href@noop {} {\bibfield  {journal} {\bibinfo  {journal} {International
  Journal of Modern Physics C}\ }\textbf {\bibinfo {volume} {17}},\ \bibinfo
  {pages} {1597} (\bibinfo {year} {2006})}\BibitemShut {NoStop}%
\bibitem [{\citenamefont {Alexopoulos}\ and\ \citenamefont
  {Leontsinis}(2014)}]{alexopoulos2014benford}%
  \BibitemOpen
  \bibfield  {author} {\bibinfo {author} {\bibfnamefont {T.}~\bibnamefont
  {Alexopoulos}}\ and\ \bibinfo {author} {\bibfnamefont {S.}~\bibnamefont
  {Leontsinis}},\ }\href@noop {} {\bibfield  {journal} {\bibinfo  {journal}
  {Journal of Astrophysics and Astronomy}\ }\textbf {\bibinfo {volume} {35}},\
  \bibinfo {pages} {639} (\bibinfo {year} {2014})}\BibitemShut {NoStop}%
\bibitem [{\citenamefont {Sambridge}\ \emph {et~al.}(2010)\citenamefont
  {Sambridge}, \citenamefont {Tkal{\v{c}}i{\'c}},\ and\ \citenamefont
  {Jackson}}]{geographical}%
  \BibitemOpen
  \bibfield  {author} {\bibinfo {author} {\bibfnamefont {M.}~\bibnamefont
  {Sambridge}}, \bibinfo {author} {\bibfnamefont {H.}~\bibnamefont
  {Tkal{\v{c}}i{\'c}}}, \ and\ \bibinfo {author} {\bibfnamefont
  {A.}~\bibnamefont {Jackson}},\ }\href@noop {} {\bibfield  {journal} {\bibinfo
   {journal} {Geophysical research letters}\ }\textbf {\bibinfo {volume} {37}}
  (\bibinfo {year} {2010})}\BibitemShut {NoStop}%
\bibitem [{\citenamefont {Busta}\ and\ \citenamefont
  {Weinberg}(1998)}]{neural}%
  \BibitemOpen
  \bibfield  {author} {\bibinfo {author} {\bibfnamefont {B.}~\bibnamefont
  {Busta}}\ and\ \bibinfo {author} {\bibfnamefont {R.}~\bibnamefont
  {Weinberg}},\ }\href@noop {} {\bibfield  {journal} {\bibinfo  {journal}
  {Managerial Auditing Journal}\ }\textbf {\bibinfo {volume} {13}},\ \bibinfo
  {pages} {356} (\bibinfo {year} {1998})}\BibitemShut {NoStop}%
\bibitem [{\citenamefont {C{\'a}ceres}\ \emph {et~al.}(2008)\citenamefont
  {C{\'a}ceres}, \citenamefont {Garc{\'\i}a}, \citenamefont
  {Mart{\'\i}nez~Ortiz},\ and\ \citenamefont {Dom{\'\i}nguez}}]{biology2}%
  \BibitemOpen
  \bibfield  {author} {\bibinfo {author} {\bibfnamefont {J.~L.~H.}\
  \bibnamefont {C{\'a}ceres}}, \bibinfo {author} {\bibfnamefont {J.~L.~P.}\
  \bibnamefont {Garc{\'\i}a}}, \bibinfo {author} {\bibfnamefont
  {C.}~\bibnamefont {Mart{\'\i}nez~Ortiz}}, \ and\ \bibinfo {author}
  {\bibfnamefont {L.~G.}\ \bibnamefont {Dom{\'\i}nguez}},\ }\href@noop {}
  {\bibfield  {journal} {\bibinfo  {journal} {Electronic Journal of
  Biomedicine}\ }\textbf {\bibinfo {volume} {1}},\ \bibinfo {pages} {27}
  (\bibinfo {year} {2008})}\BibitemShut {NoStop}%
\bibitem [{\citenamefont {Docampo}\ \emph {et~al.}(2009)\citenamefont
  {Docampo}, \citenamefont {del Mar~Trigo}, \citenamefont {Aira}, \citenamefont
  {Cabezudo},\ and\ \citenamefont {Flores-Moya}}]{aerobiological}%
  \BibitemOpen
  \bibfield  {author} {\bibinfo {author} {\bibfnamefont {S.}~\bibnamefont
  {Docampo}}, \bibinfo {author} {\bibfnamefont {M.}~\bibnamefont {del
  Mar~Trigo}}, \bibinfo {author} {\bibfnamefont {M.~J.}\ \bibnamefont {Aira}},
  \bibinfo {author} {\bibfnamefont {B.}~\bibnamefont {Cabezudo}}, \ and\
  \bibinfo {author} {\bibfnamefont {A.}~\bibnamefont {Flores-Moya}},\
  }\href@noop {} {\bibfield  {journal} {\bibinfo  {journal} {Aerobiologia}\
  }\textbf {\bibinfo {volume} {25}},\ \bibinfo {pages} {275} (\bibinfo {year}
  {2009})}\BibitemShut {NoStop}%
\bibitem [{\citenamefont {Sottili}\ \emph {et~al.}(2012)\citenamefont
  {Sottili}, \citenamefont {Palladino}, \citenamefont {Giaccio},\ and\
  \citenamefont {Messina}}]{seismic}%
  \BibitemOpen
  \bibfield  {author} {\bibinfo {author} {\bibfnamefont {G.}~\bibnamefont
  {Sottili}}, \bibinfo {author} {\bibfnamefont {D.~M.}\ \bibnamefont
  {Palladino}}, \bibinfo {author} {\bibfnamefont {B.}~\bibnamefont {Giaccio}},
  \ and\ \bibinfo {author} {\bibfnamefont {P.}~\bibnamefont {Messina}},\
  }\href@noop {} {\bibfield  {journal} {\bibinfo  {journal} {Mathematical
  Geosciences}\ }\textbf {\bibinfo {volume} {44}},\ \bibinfo {pages} {619}
  (\bibinfo {year} {2012})}\BibitemShut {NoStop}%
\bibitem [{\citenamefont {De~Ceuster}\ \emph {et~al.}(1998)\citenamefont
  {De~Ceuster}, \citenamefont {Dhaene},\ and\ \citenamefont
  {Schatteman}}]{stockmarket}%
  \BibitemOpen
  \bibfield  {author} {\bibinfo {author} {\bibfnamefont {M.~J.}\ \bibnamefont
  {De~Ceuster}}, \bibinfo {author} {\bibfnamefont {G.}~\bibnamefont {Dhaene}},
  \ and\ \bibinfo {author} {\bibfnamefont {T.}~\bibnamefont {Schatteman}},\
  }\href@noop {} {\bibfield  {journal} {\bibinfo  {journal} {Journal of
  Empirical Finance}\ }\textbf {\bibinfo {volume} {5}},\ \bibinfo {pages} {263}
  (\bibinfo {year} {1998})}\BibitemShut {NoStop}%
\bibitem [{\citenamefont {Durtschi}\ \emph {et~al.}(2004)\citenamefont
  {Durtschi}, \citenamefont {Hillison},\ and\ \citenamefont
  {Pacini}}]{financial1}%
  \BibitemOpen
  \bibfield  {author} {\bibinfo {author} {\bibfnamefont {C.}~\bibnamefont
  {Durtschi}}, \bibinfo {author} {\bibfnamefont {W.}~\bibnamefont {Hillison}},
  \ and\ \bibinfo {author} {\bibfnamefont {C.}~\bibnamefont {Pacini}},\
  }\href@noop {} {\bibfield  {journal} {\bibinfo  {journal} {Journal of
  forensic accounting}\ }\textbf {\bibinfo {volume} {5}},\ \bibinfo {pages}
  {17} (\bibinfo {year} {2004})}\BibitemShut {NoStop}%
\bibitem [{\citenamefont {Bruce~Busta}\ and\ \citenamefont
  {Sundheim}(1992)}]{taxfruad}%
  \BibitemOpen
  \bibfield  {author} {\bibinfo {author} {\bibfnamefont {C.}~\bibnamefont
  {Bruce~Busta}}\ and\ \bibinfo {author} {\bibfnamefont {R.}~\bibnamefont
  {Sundheim}},\ }\href@noop {} {\bibfield  {journal} {\bibinfo  {journal}
  {Center for Business Research}\ }\textbf {\bibinfo {volume} {95}},\ \bibinfo
  {pages} {106} (\bibinfo {year} {1992})}\BibitemShut {NoStop}%
\bibitem [{\citenamefont {Deckert}\ \emph {et~al.}(2010)\citenamefont
  {Deckert}, \citenamefont {Myagkov},\ and\ \citenamefont
  {Ordeshook}}]{electionfraud}%
  \BibitemOpen
  \bibfield  {author} {\bibinfo {author} {\bibfnamefont {J.}~\bibnamefont
  {Deckert}}, \bibinfo {author} {\bibfnamefont {M.}~\bibnamefont {Myagkov}}, \
  and\ \bibinfo {author} {\bibfnamefont {P.~C.}\ \bibnamefont {Ordeshook}},\
  }\href@noop {} {\bibfield  {journal} {\bibinfo  {journal} {Caltech/MIT Voting
  Technology Project Working Paper}\ } (\bibinfo {year} {2010})}\BibitemShut
  {NoStop}%
\bibitem [{\citenamefont {De}\ and\ \citenamefont {Sen}(2011)}]{ujjwal1}%
  \BibitemOpen
  \bibfield  {author} {\bibinfo {author} {\bibfnamefont {A.~S.}\ \bibnamefont
  {De}}\ and\ \bibinfo {author} {\bibfnamefont {U.}~\bibnamefont {Sen}},\
  }\href@noop {} {\bibfield  {journal} {\bibinfo  {journal} {EPL (Europhysics
  Letters)}\ }\textbf {\bibinfo {volume} {95}},\ \bibinfo {pages} {50008}
  (\bibinfo {year} {2011})}\BibitemShut {NoStop}%
\bibitem [{\citenamefont {Rane}\ \emph {et~al.}(2014)\citenamefont {Rane},
  \citenamefont {Mishra}, \citenamefont {Biswas}, \citenamefont {Sen},
  \citenamefont {Sen} \emph {et~al.}}]{ujjwal2}%
  \BibitemOpen
  \bibfield  {author} {\bibinfo {author} {\bibfnamefont {A.~D.}\ \bibnamefont
  {Rane}}, \bibinfo {author} {\bibfnamefont {U.}~\bibnamefont {Mishra}},
  \bibinfo {author} {\bibfnamefont {A.}~\bibnamefont {Biswas}}, \bibinfo
  {author} {\bibfnamefont {A.}~\bibnamefont {Sen}}, \bibinfo {author}
  {\bibfnamefont {U.}~\bibnamefont {Sen}},  \emph {et~al.},\ }\href@noop {}
  {\bibfield  {journal} {\bibinfo  {journal} {Physical Review E}\ }\textbf
  {\bibinfo {volume} {90}},\ \bibinfo {pages} {022144} (\bibinfo {year}
  {2014})}\BibitemShut {NoStop}%
\bibitem [{\citenamefont {Bhole}\ \emph {et~al.}(2015)\citenamefont {Bhole},
  \citenamefont {Shukla},\ and\ \citenamefont {Mahesh}}]{bhole2015benford}%
  \BibitemOpen
  \bibfield  {author} {\bibinfo {author} {\bibfnamefont {G.}~\bibnamefont
  {Bhole}}, \bibinfo {author} {\bibfnamefont {A.}~\bibnamefont {Shukla}}, \
  and\ \bibinfo {author} {\bibfnamefont {T.}~\bibnamefont {Mahesh}},\
  }\href@noop {} {\bibfield  {journal} {\bibinfo  {journal} {Chemical Physics
  Letters}\ }\textbf {\bibinfo {volume} {639}},\ \bibinfo {pages} {36}
  (\bibinfo {year} {2015})}\BibitemShut {NoStop}%
\bibitem [{\citenamefont {Shao}\ and\ \citenamefont
  {Ma}(2010)}]{shao2010significant}%
  \BibitemOpen
  \bibfield  {author} {\bibinfo {author} {\bibfnamefont {L.}~\bibnamefont
  {Shao}}\ and\ \bibinfo {author} {\bibfnamefont {B.-Q.}\ \bibnamefont {Ma}},\
  }\href@noop {} {\bibfield  {journal} {\bibinfo  {journal} {Physica A:
  Statistical Mechanics and its Applications}\ }\textbf {\bibinfo {volume}
  {389}},\ \bibinfo {pages} {3109} (\bibinfo {year} {2010})}\BibitemShut
  {NoStop}%
\bibitem [{\citenamefont {Lütkebohle}(2016)}]{keplerarchive}%
  \BibitemOpen
  \bibfield  {author} {\bibinfo {author} {\bibfnamefont {I.}~\bibnamefont
  {Lütkebohle}},\ }\href@noop {} {\enquote {\bibinfo {title} {Exoplanet data
  archeive for confirm planets},}\ }\bibinfo {howpublished}
  {\url{http://exoplanetarchive.ipac.caltech.edu/cgi-bin/TblView/nph-tblView?app=ExoTbls&config=planets}}
  (\bibinfo {year} {2016})\BibitemShut {NoStop}%
\bibitem [{\citenamefont {Pintr}\ \emph {et~al.}(2013)\citenamefont {Pintr},
  \citenamefont {Pe{\v{r}}inov{\'a}}, \citenamefont {Luk{\v{s}}},\ and\
  \citenamefont {Pathak}}]{pintr2013statistical}%
  \BibitemOpen
  \bibfield  {author} {\bibinfo {author} {\bibfnamefont {P.}~\bibnamefont
  {Pintr}}, \bibinfo {author} {\bibfnamefont {V.}~\bibnamefont
  {Pe{\v{r}}inov{\'a}}}, \bibinfo {author} {\bibfnamefont {A.}~\bibnamefont
  {Luk{\v{s}}}}, \ and\ \bibinfo {author} {\bibfnamefont {A.}~\bibnamefont
  {Pathak}},\ }\href@noop {} {\bibfield  {journal} {\bibinfo  {journal}
  {Planetary and Space Science}\ }\textbf {\bibinfo {volume} {75}},\ \bibinfo
  {pages} {37} (\bibinfo {year} {2013})}\BibitemShut {NoStop}%
\bibitem [{\citenamefont {Pintr}\ \emph {et~al.}(2014)\citenamefont {Pintr},
  \citenamefont {Pe{\v{r}}inov{\'a}}, \citenamefont {Luk{\v{s}}},\ and\
  \citenamefont {Pathak}}]{pintr2014relative}%
  \BibitemOpen
  \bibfield  {author} {\bibinfo {author} {\bibfnamefont {P.}~\bibnamefont
  {Pintr}}, \bibinfo {author} {\bibfnamefont {V.}~\bibnamefont
  {Pe{\v{r}}inov{\'a}}}, \bibinfo {author} {\bibfnamefont {A.}~\bibnamefont
  {Luk{\v{s}}}}, \ and\ \bibinfo {author} {\bibfnamefont {A.}~\bibnamefont
  {Pathak}},\ }\href@noop {} {\bibfield  {journal} {\bibinfo  {journal}
  {Planetary and Space Science}\ }\textbf {\bibinfo {volume} {99}},\ \bibinfo
  {pages} {1} (\bibinfo {year} {2014})}\BibitemShut {NoStop}%
\bibitem [{nat(2014)}]{natureasia}%
  \BibitemOpen
  \href@noop {} {\enquote {\bibinfo {title} {Looking for habitable planets
  beyond solar system},}\ }\bibinfo {howpublished}
  {\url{http://www.natureasia.com/en/nindia/article/10.1038/nindia.2014.123}}
  (\bibinfo {year} {2014})\BibitemShut {NoStop}%
\bibitem [{\citenamefont {Lütkebohle}()}]{benfordonexcel}%
  \BibitemOpen
  \bibfield  {author} {\bibinfo {author} {\bibfnamefont {I.}~\bibnamefont
  {Lütkebohle}},\ }\href@noop {} {\enquote {\bibinfo {title} {Step by step
  instructions for using benford law},}\ }\bibinfo {howpublished}
  {\url{http://www.theiia.org/intAuditor/media/files/Step-by-step_Instructions_for_Using_Benford's_Law[1].pdf}}\BibitemShut
  {NoStop}%
\bibitem [{\citenamefont {Bhole}\ \emph {et~al.}(2014)\citenamefont {Bhole},
  \citenamefont {Shukla},\ and\ \citenamefont {Mahesh}}]{gaurav}%
  \BibitemOpen
  \bibfield  {author} {\bibinfo {author} {\bibfnamefont {G.}~\bibnamefont
  {Bhole}}, \bibinfo {author} {\bibfnamefont {A.}~\bibnamefont {Shukla}}, \
  and\ \bibinfo {author} {\bibfnamefont {T.}~\bibnamefont {Mahesh}},\
  }\href@noop {} {\bibfield  {journal} {\bibinfo  {journal} {arXiv preprint
  arXiv:1406.7077}\ } (\bibinfo {year} {2014})}\BibitemShut {NoStop}%
\bibitem [{\citenamefont {Hill}(1995)}]{hill1995base}%
  \BibitemOpen
  \bibfield  {author} {\bibinfo {author} {\bibfnamefont {T.~P.}\ \bibnamefont
  {Hill}},\ }\href@noop {} {\bibfield  {journal} {\bibinfo  {journal}
  {Proceedings of the American Mathematical Society}\ }\textbf {\bibinfo
  {volume} {123}},\ \bibinfo {pages} {887} (\bibinfo {year}
  {1995})}\BibitemShut {NoStop}%
\bibitem [{wik()}]{wikiefftemp}%
  \BibitemOpen
  \href@noop {} {\enquote {\bibinfo {title} {Effective teamperature},}\
  }\bibinfo {howpublished}
  {\url{https://en.wikipedia.org/wiki/Effective_temperature}}\BibitemShut
  {NoStop}%
\bibitem [{\citenamefont {Miller}\ \emph {et~al.}(2015)\citenamefont {Miller},
  \citenamefont {Berger},\ and\ \citenamefont {Hill}}]{miller2015theory}%
  \BibitemOpen
  \bibfield  {author} {\bibinfo {author} {\bibfnamefont {S.~J.}\ \bibnamefont
  {Miller}}, \bibinfo {author} {\bibfnamefont {A.}~\bibnamefont {Berger}}, \
  and\ \bibinfo {author} {\bibfnamefont {T.}~\bibnamefont {Hill}},\ }\href@noop
  {} {\enquote {\bibinfo {title} {The theory and applications of benford’s
  law},}\ } (\bibinfo {year} {2015})\BibitemShut {NoStop}%
\end{thebibliography}%

\end{document}